\documentclass{mem}
\usepackage{natbib}\usepackage{txfonts}
\usepackage{balance}
\usepackage{graphicx}
\idline{0}{1}
\begin{document}
\def\teff{$T\rm_{eff }$}
\def\kms{$\mathrm {km s}^{-1}$}

\title{3D stellar atmospheres 
for stellar structure models and asteroseismology}

   \subtitle{}

\author{
F. \,Kupka\inst{1,2} 
          }

  \offprints{F. Kupka}

\institute{
Observatoire de Paris, LESIA, CNRS UMR 8109, 
F-92195 Meudon, France
\and
now at Institute of Mathematics, University of Vienna, Nordbergstra{\ss}e 15,
A-1090 Vienna, Austria\quad
\email{Friedrich.Kupka@univie.ac.at}
}

\authorrunning{Kupka}

\titlerunning{3D Stellar Atmospheres, Stellar Structure, and Asteroseismology}

\abstract{
Convection is the most important physical process that determines the structure of 
the envelopes of cool stars. It influences the surface radiation flux and the shape of
observed spectral line profiles and is responsible for both generating and damping solar-like
oscillations, among others. 3D numerical simulations of stellar surface convection
have developed into a powerful tool to model and analyse the physical mechanisms
operating at the surface of cool stars. This review discusses the main principles of
3D stellar atmospheres used for such applications. The requirements from
stellar structure and evolution theory to use them as boundary conditions are
analysed as well as the capabilities of using helio- and asteroseismology
to reduce modelling uncertainties and probing the consistency and accuracy
of 3D stellar atmospheres as part of this process. Simulations for the solar surface
made by different teams  are compared and some issues concerning the
uncertainties of this modelling approach are discussed. 
}
\maketitle{}

\section{Introduction -- 3D Stellar Atmospheres}

The inhomogeneous solar surface is characterised by large scale structures
originating from the solar convection zone as well as from the complex solar
magnetic field. Globally, the solar convection zone is dominated by an extreme
density and temperature stratification with a contrast of $625,000:1$ and $350:1$,
respectively, when comparing values for these quantities as found for the bottom of
the solar convection zone with those found for its surface. The model data underlying
these estimates are supported by helioseismology and solar surface observations
(for further references related to these estimates see \citealt{Kupka09}). As a result
the entire solar convection zone comprises 20 pressure scale heights spread over 
30\% of the solar radius. In spite of the slow solar rotation rate the timescales on
which convection transports mass and heat through the entire zone are hence 
comparable to those of rotation. This does not hold for the surface layers of the
Sun. With average velocities reaching 2 to 3~km~s$^{-1}$ the granules and
downflow regions observed at the solar surface change over timescales of
a few minutes to less than half an hour, two orders of magnitudes faster than
the timescale of rotation on these length scales. Similar holds for other types
of stars with convection zones reaching the stellar surface provided that the
surface gravity is high enough such that the pressure scale height at
optical depths of order unity is just a small fraction of the stellar radius.

These physical relations are the main background for the ``box-in-a-star'' 
approach to 3D stellar atmospheres. In this case, a small volume of the entire
convection zone is considered for which the conservation law equations for the
densities of mass, momentum, and energy, have to be solved. The hydrodynamical
equations are coupled to the equation for radiative transfer, which have to be solved
as well, since at the surface of a star the diffusion approximation is no longer valid.
Realistic hydrodynamical simulations of such a system require a tabulated equation
of state and tabulated opacities computed for the chemical composition assumed
for the object to be studied. The simulations performed for this setting describe the 
evolution of such a sample volume of the stellar atmosphere (and upper envelope)
in space and time, as represented on a grid extending over the entire simulation 
box and computed for a finite number of time steps. The capabilities of this approach
and its successful recovery of observational data such as surface intensity images in 
the visual or detailed spectral line profiles have been discussed in many research papers
and conference reviews (cf.\ \citealt{Stein98} as well as examples and references
in \citealt{Asplund07}, \citealt{Steffen07}, \citealt{Kupka09}, and the extensive
review of \citealt{Nordlund09}). 

In the following we provide a short review on how 3D stellar atmospheres can
be used for the modelling of stellar structure and for helio- and asteroseismology.
We discuss requirements on model stellar atmospheres posed by these branches
of stellar physics and some applications to illustrate the demands on the
3D models. More recently, 3D stellar atmospheres have been used to
compute physical quantities difficult to access by means of direct observations
or even seismology, such as turbulent pressure or skewness of vertical
velocities below the visible solar or stellar surface. We provide some
first results from a comparison of different 3D stellar atmospheres used to 
compute statistically averaged quantities required in stellar structure modelling
and seismology. We also discuss some issues related to the resolution
of 3D models and conclude this paper with a summary.

\section{Stellar Modelling}

A solar structure model or a stellar evolution track depends on a variety
of hydrodynamical processes including convection. Since the associated
hydrodynamical timescales are many orders of magnitudes shorter
than the thermal and nuclear ones during most phases of stellar
evolution, 3D hydrodynamical simulations of solar or stellar evolution 
are computationally too demanding. This restriction holds
even for simulations of solar convection on just the hydrodynamical timescales
because of extreme stratification (20 pressure
scale heights from the top to the bottom of the zone). 

From a stellar structure modelling point of view convection models are
required, among others, to compute stellar radii of stars which have convective
envelopes. For this purpose tables of the mixing length parameter and
turbulent pressure obtained from hydrodynamical simulations of the surface
convection zones would in principle be sufficient (see \citealt{Ludwig98}, 
\citealt{Trampedach99}, and more recent work from the same authors for
some examples). But for stellar evolution modelling such tables are inadequate. 
In particular, these tables would not describe the nature of convection in the
stellar interior, where we need to compute
the amount overshooting of convective zones into radiative ones,
the precise convective efficiency in those transition regions, and other
phenomena which occur in stellar interiors such as semi-convection and 
further varieties of double-diffusive convection. The currently used
analytical, semi-empirical formulae for the description of these
physical processes suffer from inconsistencies (cf.\ the discussions
in \citealt{CM91,Canuto93,Charb07}). Hence,
the development of improved calibration methods is highly rewarding
as are more refined simulations, which would be suitable for
deducing more versatile convection models.

Classical procedures for the computation of stellar models with convective
envelopes such as calibrating the mixing length have some severe limitations.
Among others, they are based on integral quantities (radius, luminosity, etc.). 
Unless secondary calibration methods are used (which in turn require stellar
models), these quantities are known only for rather few, nearby stars with 
sufficient accuracy. In the end, such methods are an ambiguous probe of 
convection models.

Methods based on detailed photospheric properties alone (such as 
comparing observed with computed spectral line profiles) require 
quantities such as the mean temperature to be properly 
modelled as a function of depth. They hence provide some extra
information to reduce ambiguities in calibrations and models tests,
but they have only limited implications on how convection is modelled
in the stellar interior and can still remain inconclusive for photospheric
convection modelling (cf.\ \citealt{Montalban04} and \citealt{Heiter02},
for example).

\section{Helioseismology and Asteroseismology}

Helio- and asteroseismology have opened a new window to look inside
stars. They allow a much more thorough understanding of the mean
structure and hydrodynamical properties of stellar interiors due to
convection and rotation. Traditional ``seismological tests'' of
convection models are based on frequency information obtained
from helioseismological or asteroseismological observations. But in
spite of the constraints they provide on measurements of the
depth of the solar convection zone, on the extent of deviations
from a radiative temperature gradient in the layers right
underneath the solar convection zone (overshooting), or on the
chemical composition, these methods are still limited by ambiguities.
One of the reasons is that the acoustic size of
a resonant cavity can be the same for different model structures
within the frequency domain which is accessible to measurements
with sufficient accuracy. This is the physical background for two
contradicting explanations suggested for the differences
between observed solar p-mode frequencies and predicted ones
based on standard solar models (\citealt{Basu95} found that a much
steeper superadiabatic temperature gradient as obtained with
the convection model by \citealt{CM91,CM92} reduces the
discrepancy between observations and computations, while
a similar reduction was found when using averaged model structures
from 3D hydrodynamical simulations of the surface layers by
\citealt{Rosenthal99}, who pointed out the role of turbulent pressure
and differences in the model structure originating from the 3D,
inhomogeneous radiative transfer --- see also \citealt{CCD98}
for further discussion).

Observational data on amplitude and time dependence (mode 
linewidth and lifetime) of the p-modes, i.e.\ the study of p-mode excitation and
p-mode damping, provides the extra information necessary to remove
such ambiguities. Mode excitation occurs due to shear stresses
and entropy fluctuations. Based on earlier work by 
\citet{Gold77} and \citet{Balmforth92}, different 
semi-analytical models have been developed (see
\citealt{Houdek99,Samadi01,Houdek02, Chaplin05}). 
Through predictions of velocity and
convective flux these models link observations with 
quantities required to construct stellar models. Models
or numerical simulations are used to
compute eigenfunctions and associated eigenfrequencies,
the mean structure (density, entropy, \dots), spatial and temporal
correlations of velocity and entropy as a function of length scale, 
the filling factor (fraction of horizontal area covered by upflows),
and the ratio of vertical to horizontal root mean square velocities.

Observations (with some modelling) provide
quantities such as mode mass, the height above the photosphere where
the modes are measured, the mode line width at half maximum, and the
mean square of the mode surface velocity. \citet{Samadi06} have used
standard solar structure models to compute the model dependent input
data for the approach of \citet{Samadi01} required to calculate p-mode
excitation rates. They investigated the influence of non-grey atmospheres in
comparison with the grey diffusion approximation and compared a standard 
mixing length model of convection with the model of \citet{CGM96}.
The standard model (grey, mixing length treatment of convection) 
underestimates excitation rates by up to an order of magnitude.
Convection treatment was found to be more important for the 
excitation rate calculation than line blanketing (non-grey atmospheres),
but eventually also the convection model by \citet{CGM96} combined
with a non-grey atmosphere, as in \citet{Heiter02}, was found to yield
a model structure that still underestimates p-mode excitation rates
by up to a factor of 3.

By comparison, numerical simulations were found to reproduce
observed excitation rates under the assumption of a non-Gaussian
temporal correlation \citep{Samadi03}. A new, semi-analytical model 
to compute p-mode excitation rates was presented by \citet{Belkacem06a}.
It is based on the model of \citet{Gryanik02} and \citet{Gryanik05}
for computing third and fourth order correlation functions in flows
dominated by turbulent convection. This model accounts for
asymmetry between up-  and downflows (skewness of velocity and
temperature). Due to mass conservation a flow with broad upflows
(granules !) has higher velocities in its downflows and since the input
power for p-modes depends quadratically on the velocity, the total
input power increases in this case. \citet{Belkacem06b} have shown
that this model together with the assumption of a Lorentzian
distribution function taken for the temporal correlations 
predicts solar p-mode excitation rates in agreement with observations.
We note at this point that some of the input for their model had to
be taken from numerical simulations (for instance, the horizontal
area fraction covered by upflow regions). 

The applicability of their new approach to stars other than the
Sun was shown by \citet{Samadi08} by computing mode excitation
rates for $\alpha$~Cen~A and comparing them with observational
predictions. The model predictions were found in agreement with
the data, but it was also pointed out that a higher accuracy of
astero\-seismological measurements is still needed to provide
sufficiently tight constraints for the models, which is now
possible with the results from the CoRoT mission 
(\citealt{Appourchaux08,Michel08}) and soon by
the Kepler mission. This demonstrates that helio- and asteroseismology
have the capability to falsify models and simulations where classical
methods remain ambiguous. 

\section{Comparing Simulations, Resolution Issues}

How well do the currently available numerical simulations of
surface convection agree about the mean structure of the uppermost
layers of stars and on the correlation functions required to interpret
data from helio- and asteroseismology? To find an answer one should
consider, among others, the following questions: to what extent is the
flow influenced by the boundary conditions? How large is the influence 
of the domain size? How important is non-grey radiative transfer 
for the structure of the superadiabatic layer developed by convection
zones that reach the stellar surface? Do different numerical methods
and viscosity models change the large scale (coherently structured) 
part of the flow? 

In the following we discuss some first results of a comparison of solar
surface convection simulations performed with four different simulation
codes. Each of these 3D simulations is set up for the ``box-in-a-star''
scenario with a rectangular box covering only a small fraction of the 
stellar surface layers, with a constant surface gravity, and with periodic boundaries
along the horizontal directions. Results for the CO$^5$BOLD \citep{Freytag02}
have kindly been provided by M.~Steffen. This code uses a Roe-scheme
for advection (i.e.\ a non-linear numerical viscosity) combined with
a subgrid-scale viscosity \citep{Smagorinsky63}. For the two simulations shown
the vertical boundaries are open and the chemical composition assumed is
that one of \citet{G98}. The high resolution case shown assumes
non-grey radiative transfer (with 5 opacity bins), a box volume of 
$11.2^2 \times 3.1$~Mm$^3$, a grid of $400^2*165$ points, a constant
horizontal grid size of $28$~km and a variable vertical one of $12\dots 28$~km.
The deep simulation assumes grey radiative transfer, a box volume of 
$11.2^2 \times 5.2$~Mm$^3$, a grid of $200^2\times 250$ points, a constant
horizontal grid size of $56$~km and a fixed vertical one of $21$~km.

Two further simulations have kindly been provided by F.J.~Robinson
and are based on the simulation code presented in \citet{KC98} and
\citet{Robinson03} (named CKS code in the figures). Its numerical scheme uses plain
second order finite differences with a subgrid-scale viscosity \citep{Smagorinsky63}
where the coefficients are boosted near shock fronts. Both sets have closed vertical
boundaries and grey radiative transfer is assumed, with a 3D Eddington approximation.
The simulation with smaller domain size (``case D'' from \citealt{Robinson03})
assumes a chemical composition as in \citet{G93} and has a box volume of 
$2.9^2 \times 3$~Mm$^3$, a grid size of $58 ^2\times 170$ points, a constant
horizontal grid size of $50$~km and a fixed vertical one of $17.6$~km. The
``2008'' simulation assumes the chemical composition of \citet{G98} and
has a box volume of $5.4^2 \times 3.6$~Mm$^3$, a grid of 
$117^2\times 190$ points, a constant horizontal grid size of $46$~km 
and a fixed vertical one of $19$~km. 

Another simulation, performed with the code by \citet{Stein98}, has
kindly been provided by R.~Samadi and K.~Belkacem (it is essentially the
one used by \citealt{Belkacem06a}). The advection scheme of this code 
is stabilized with hyperviscosity and an extra viscosity near shock fronts.
Open vertical boundary conditions are assumed and for this simulation
the chemical composition is similar to that one published by \citet{G98}.
A non-grey radiative transfer (with 4 bins) is assumed as well as a box
volume of $6^2 \times 3$~Mm$^3$, a grid of $150^3$ points,
a constant horizontal grid size of $40$~km and a variable vertical grid 
with an average spacing of $20$~km. 

Finally, we present results from a new simulation performed with the
ANTARES code \citep{Muthsam07,Muthsam09}. Due to its high order 
advection scheme based on essentially non-oscillatory methods, i.e.\
a non-linear numerical viscosity, physical viscosity sources (radiative
and molecular ones) are sufficient to stabilize the code for the
photospheric layers contained in the simulation. For the stellar interior
their contributions are negligible and the numerical scheme is stable
without adding subgrid-scale or artificial diffusivities. For this
simulation the chemical composition presented in \citet{G93} is
used together with a box volume of $6^2 \times 3$~Mm$^3$, a grid
of $150^2\times 190$ points, a constant horizontal grid size of 
$40$~km and a constant vertical grid spacing of $16$~km, and 
grey radiative transfer is assumed. 

\begin{figure*}[t!]
\resizebox{\hsize}{!}{\includegraphics[width=0.6\textwidth]{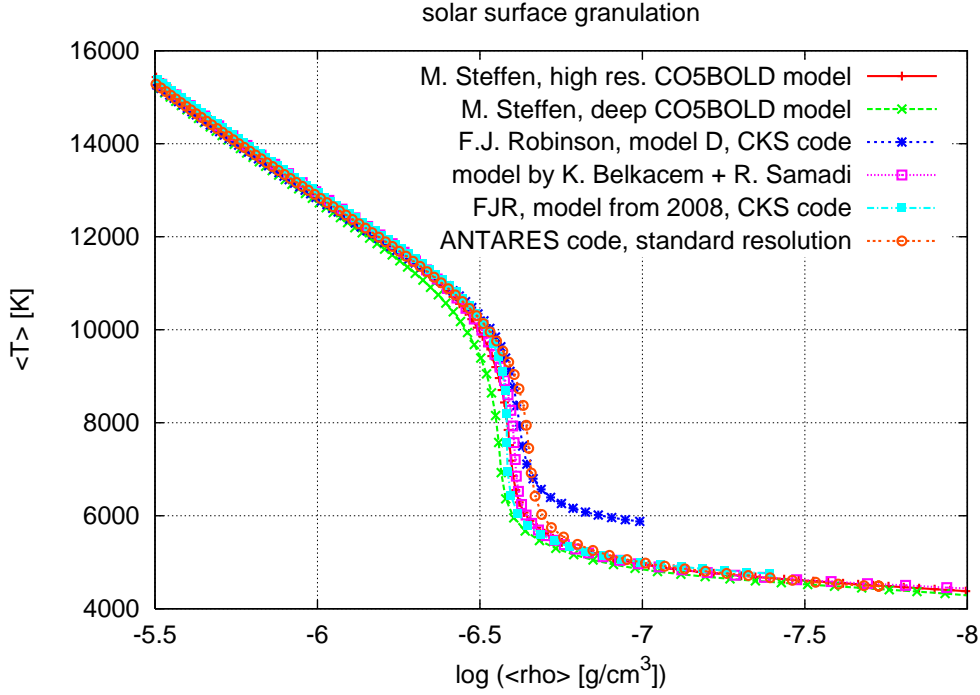}}
\caption{
\footnotesize
Horizontally averaged mean temperature $T$ as a function of the logarithm
of the horizontally averaged mean density $\rho$ for the six simulation sets.
The region shown includes the uppermost layers of the solar interior, the
superadiabatic peak, and the lower photosphere.
}
\label{Fig_FK_1}
\end{figure*}

In Fig.~\ref{Fig_FK_1} we compare the horizontally averaged mean 
temperature $T$, plotted as a function of the logarithm of the horizontally 
averaged mean density $\rho$, for all six simulation sets. The averaging
was performed for each horizontal layer and each point in time 
followed by a time average. Since there is no inversion of the mean
density as a function of depth in any of the simulations, there is a unique
mapping of the mean density to depth and radius for each simulation. 
In the interior all simulations agree very well with each other. Near the
superadiabatic peak the maximum spread in mean density for a given
mean temperature is $\pm 15\%$ and is due to different metallicities,
different resolution, and different treatment of radiative transfer schemes
being used. For the lower photosphere all but one of the simulations agree
again quite well (the exception is the ``case D'' model which does not
extend far enough into the photosphere to produce realistic temperatures
in that region). For the mid and upper photosphere, which are not shown
here, the simulations with grey radiative transfer yield higher average
temperatures, although the detailed behaviour for the outermost layers
also depends on the exact implementation of the boundary conditions.

\begin{figure*}[t!]
\resizebox{\hsize}{!}{\includegraphics[]{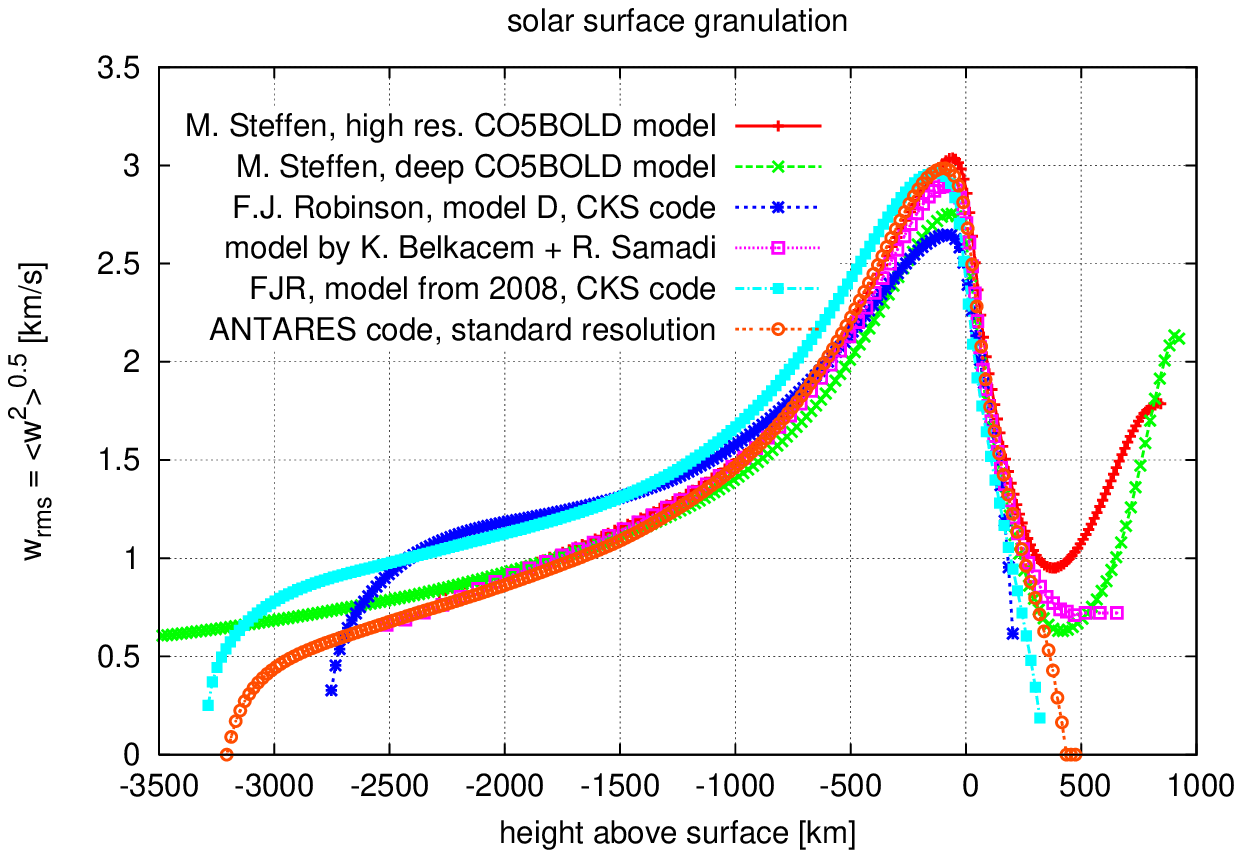}}
\caption{
\footnotesize
Root mean square values of horizontally averaged vertical velocity
fluctuations of all six simulations.
}
\label{Fig_FK_2}
\end{figure*}

In Fig.~\ref{Fig_FK_2} we compare the root mean square values
of the horizontally averaged vertical velocity fluctuations of all six
simulations. Despite the simulations with closed boundary conditions have
$w_{\rm rms}$ values which tend towards zero near the top, they 
behave very similarly to the simulations with open vertical boundaries
for the layers at the solar surface up to the highest velocities, which
are found around the superadiabatic peak. Only the shallow ``model D''
computed with the CKS code and the lower resolution, deep model
computed with the CO$^5$BOLD code predict slightly lower maximum
values. Further inside, both simulations with the CKS code predict
higher values for $w_{\rm rms}$, which drops rapidly near the closed
bottom. In turn, the averages from the ANTARES code are very 
similar to those computed by K.~Belkacem and R.~Samadi with
the code by \citet{Stein98} from the solar surface close to the
bottom of the simulation volume. Note the different behaviour of
the simulation codes using open vertical boundaries near the top
of the simulation domain (increasing vs.\ constant $w_{\rm rms}$). 

\begin{figure*}[t!]
\resizebox{\hsize}{!}{\includegraphics[]{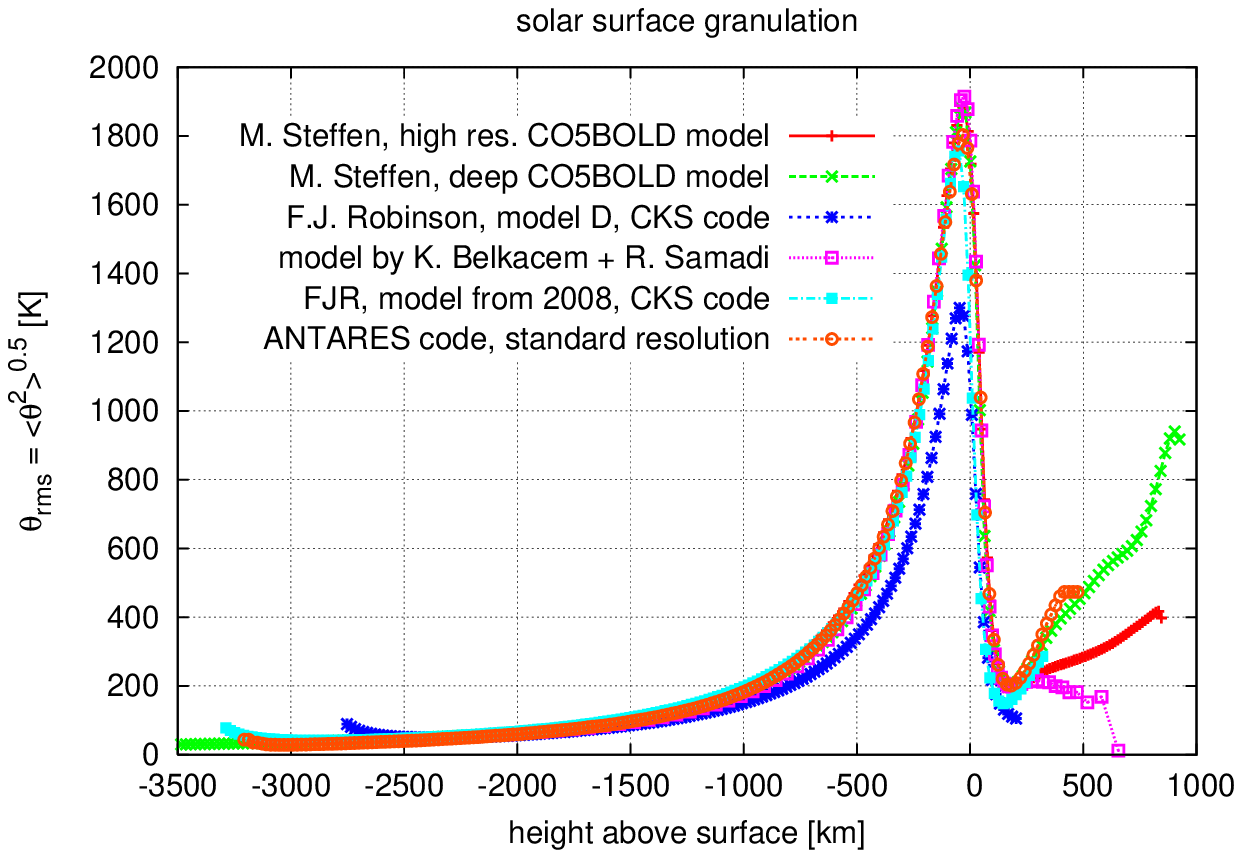}}
\caption{
\footnotesize
Root mean square values of the horizontally averaged temperature
fluctuations of all six simulations.
}
\label{Fig_FK_3}
\end{figure*}

In Fig.~\ref{Fig_FK_3} we compare the root mean square values
of the horizontally averaged temperature fluctuations of all six simulations.
Differences in the mid and upper photosphere originate from the non-grey
radiative transfer assumed in some of the simulations. Except for the
most shallow model, which has lower peak values, the root mean square
fluctuations agree well among all of the simulations throughout most of the
convection zone. Simulations with non-grey radiative transfer have larger
peak values by only a few percent. Simulations with closed lower vertical
boundary can be distinguished by an increase of temperature fluctuations
close to the bottom boundary layer.

\begin{figure*}[t!]
\resizebox{\hsize}{!}{\includegraphics[]{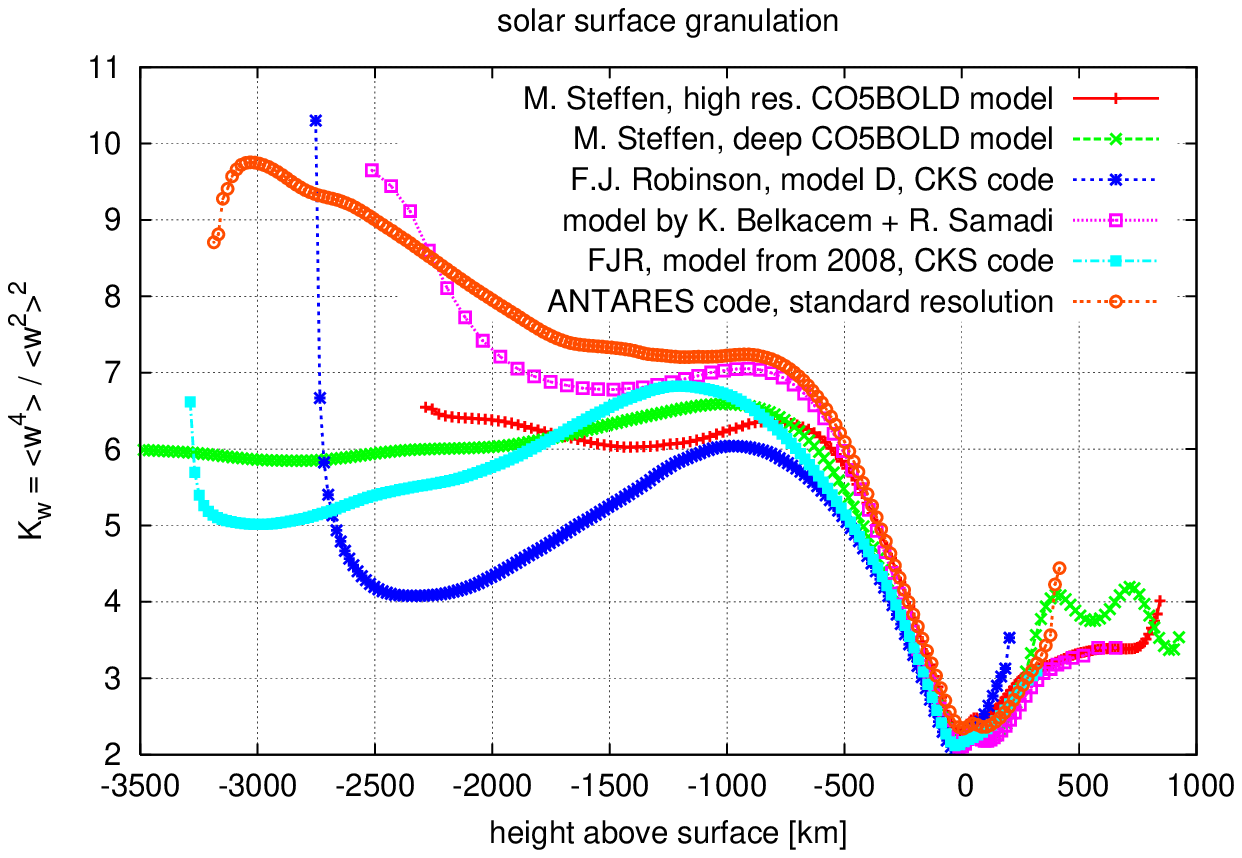}}
\caption{
\footnotesize
Kurtosis $K_w$ of horizontally averaged vertical velocity fluctuations
for all six simulations.
}
\label{Fig_FK_4}
\end{figure*}

In Fig.~\ref{Fig_FK_4} we compare the kurtosis $K_w$ of the horizontally 
averaged vertical velocity fluctuations for all six simulations. We note
that again all simulations agree for the layers from the bottom of the
photosphere till below the superadiabatic peak. However, further below
the differences become large despite long-term averages performed
over about 4 to 12 hours of solar time (about 15 to 50 convective turnover
times) have been used. In particular, simulations with open vertical
boundaries {\em do not (!)} agree more closely among each other when  
compared to simulations with closed vertical boundaries. This is an 
important caveat to remember when using 3D stellar atmospheres
to compute data for helio- and asteroseismology. Clearly, the higher
order correlations are much more sensitive to the details of the
implementation of the boundary conditions than the variables 
representing only the mean structure of the simulation domains.

Since the simulations compared above have the same resolution
within a factor of two, one might argue how they depend on numerical
methods used and on resolution in general. None of the
simulations compared above can be expected to resolve turbulence
created by shear between up- and downflows, since in 3D one 
expects a horizontal and vertical resolution of less than 5~km to
be required for that \citep{Kupka09}. Questions related to resolution 
include whether the spectral line profiles remain unaffected
by turbulence generated through shear underneath the visible surface,
what are the contributions of turbulence to Reynolds stresses and entropy
for p-mode driving, or whether a better resolved radiative cooling
at the stellar surface does change the flow properties. Resolution
effects in 3D stellar atmosphere for grid spacings similar to the
simulations discussed above have been discussed by \citet{Stein98}
and \citet{Robinson03}, among others. Recently, \citet{Muthsam09} have
demonstrated that for this resolution the diffusion of the advection scheme
in the simulation clearly has an influence on the small scale structures
at the visible surface. However, horizontal resolutions of 10~km with
the least dissipative numerical scheme are necessary to finally reveal
the highly turbulent nature of downdrafts underneath the solar surface
rather than just noticing indications for their turbulent behaviour.
At the transition between laminar and turbulent flow the effective 
resolution of a numerical scheme is crucial, since only for even higher
resolution (i.e.\ in the turbulent regime) such numerical details can be
expected to no longer matter. The influence of shear driven turbulence
on observable quantities has yet to be investigated.

\section{Summary}

Hydrodynamical simulations have developed into a tool for the
computation of realistic 3D stellar atmospheres of cool stars.
Particularly for stars at or close to the main sequence they have
successfully passed observational tests based on spectroscopy
and traditional calibration methods for convection models, 
but also tests based on helioseismology, which can distinguish much
more sensitively between different physical models of the outer envelope
of the Sun. Asteroseismology is now gradually gaining the same
capability. Current numerical simulations of solar surface convection 
are robust in their predictions of the mean solar surface structure 
(different boundary conditions, radiative transfer treatment, resolution) 
and pass observational tests which require moderate spatial resolution.
However, a comparison of higher order correlation functions for velocity
(and temperature) demonstrates that these quantities are sensitive to
the implementation of boundary conditions. Interestingly, they could
be tested by means of helio- and asteroseismology.
Refined advection numerics and high spatial resolution show how
turbulence is generated in downflows shrouded by the surface
layers (cf. also \citealt{Stein2000} for a first study of this question).

Thus, 3D stellar atmospheres provide stellar evolution
modelling with a tool to calibrate convection models and 
improve stellar evolution calculations. Both spectroscopy and
asteroseismology will continue to help in defining the region of
applicability of numerical simulations of stellar convection.
At the same time, questions related to mixing due to convection zones
and their behaviour deep inside the stars remain open.
A convection model taking these processes into account
is still in demand.

\balance 
\begin{acknowledgements}
I am grateful to the IAU for a travel grant and to the CoRoT team
at LESIA at Obs. de Paris-Meudon, in particular to M.-J.~Goupil, for
funding my participation at the IAU GA in Rio de Janeiro and its JD~10,
and to the Austrian Science Foundation FWF for support through
project P21742 while writing this contribution.
I am indebted to K.~Belkacem, F.J.~Robinson, R.~Samadi,
and M.~Steffen who have provided me with data from their
numerical simulations. Contributions by the group of
H.J.~Muthsam at Univ.~of Vienna, F.~Zaussinger at MPI
for Astrophysics, Garching, and J.~Ballot at the 
Lab.~d'Astrophysique in Toulouse to the simulations with
the ANTARES code presented in this review are gratefully
acknowledged.
\end{acknowledgements}
\bibliographystyle{aa}

\end{document}